\documentstyle[preprint]{ptptex}

\preprintnumber[3cm]{KNK-02I1}

\title{
 On the Chern-Simons Gauge Theory of Anyons \\ in the Fractional Quantum Hall Effect
}

\author{
Hitoshi {\sc Ito}\footnote{
 E-mail address: itoh@phys.kindai.ac.jp}
}

\inst{
Department of Physics, Faculty of Science and
 Engineering,\\ Kinki University, Higashi-Osaka, 577-8502, Japan \\
                January 9, 2002}

\date{}

\abst{
This short note was born out of discussions on anyons in the FQHE at the YITP workshop ``Fundamental problems of quantum field theories" (December 19-21, 2001, YITP, Kyoto). At that time, I felt that there might not be a sound consensus of opinion on the subject. Now, I would like to show my understanding here, the essential part of which is based on Ref.\cite{Ito02}.

The first problem discussed is a notion of ``bosonized electrons (bosonization)", which are unphysical objects from the standpoint of the Chern-Simons gauge invariance. Therefore, their condensate is merely of mathematical concept and the true physical state realized is a liquid-like one made of degeneration of anyons. Based on this recognition, I argue about a mechanism of the degeneration that results from the genuine CS gauge field theory. It is noted that the Ginzburg-Landau effective theory is not necessary. As the last problem, the gauge invariance in the ``composite-fermion theories" is discussed.
}

\begin{document}

\maketitle

If we want to describe a FQHS in terms of the many-body quantum mechanics, the best wave function is that of Laughlin's. The basic characteristics of it are as follows: (A) It represents a uniformly extended distribution of electrons. (B) All its zeros are located at the positions of an electron and they indicate short-range repulsive interactions between two electrons. Now, the theory must be converted into some field theory with CS interactions included. The CS field theory is a gauge theory that is specified with a background field and a statistical parameter $\alpha$ which is determined though physical conditions.\cite{Ito02} We consider two varieties of the theory. In the first one called ``background-boson (BB) gauge", the background field is assumed to be bosonic and in the second ``background-fermion (BF) gauge" it is assumed to be fermionic. Since the theory is a gauge theory, physical quantities must be gauge independent and therefore we can use any gauge in order to obtain them. In this respect, it is important to recognize correctly the basic physical fields(quasiparticles). Generally speaking, the only physical fields of the theory are anyon fields that are obtained by eliminating the CS fields by applying singular gauge transformations.

If we choose the BF gauge and assume that $\alpha=0$, the fictitious CS field decouples and the BF is nothing but an electron itself. However, this gauge is not useful because we cannot obtain the Laughlin state but can at most verify it by numerical calculations in this gauge. The most suitable gauge for investigation of the FQHE is the BB gauge. The both characteristics (A) and (B) are realized in this gauge. When an external magnetic field is applied, the ground state of the BB field becomes static and homogeneous and is regarded as the Bose condensate, that is a realization of (A). This condensate is, however, a mathematical artifact made of the fictitious bose field which couples with the CS field. Before the condensation occurs, the configulation of the boson field can be regarded as a vortex, which is accompanied by the CS field. Then, if we eliminate the CS field and attach the CS fluxes to the vortex by a gauge transformation, the vortex is converted into a physical anyon. A system of anyons thus constructed is controlled by the anyonic exclusion principle and we have just shown that the ground state of it is the uniform distribution of them represented by the condensate of the BB field. Therefore, the precise word to express the ground state is not ``condensation" but ``degeneration" under the exclusion principle. The characteristic (B) of the Laughlin wave function reflects this principle. This is the scenario by which Heaven creates the Laughlin state.

The idea of the condensation of bosons(bosonized electrons) was first investigated with the assumption that the bosons undergo changes produced by unknown self-interaction through a Higgs-type potential and the whole process of the condensation is governed by a Ginzburg-Landau (GL) effective field theory.\cite{CSGL} The bosonizing transformation means a transformation which removes the phases of the Laughlin wave function. It is, therefore, the transformation which transforms anyons back into the background boson in our terminology. An important progress was made by Ezawa, Hotta and Iwazaki, when they found that the self-interactions other than the Coulomb repulsion are not necessary for the condensation with an energy gap and for vortex excitations to occur in it.\cite{Ezawa91} We rely on their results and do not assume any Higgs potential. The theory thus constructed is the genuine CS field theory for anyons.

The last subject to be discussed is the gauge independence in the BF gauge. In the previous paper,\cite{Ito02} we have pointed out that the composite fermion is not a physical object and the real reason of disappearance of the interaction for a filling fraction $\nu=1/2$ is not a cancellation of the external magnetic field by the CS field. The real thing occured is a neutralization of the fermion. The gauge independence of results of calculations is sometimes difficult to be verified in the BF gauge. We, however, note that the correct gauge-independent value 0 of the charge of an anyon appearing in the state $\nu=1/2$ can be obtained also in the BF gauge.\cite{Murthy98}


\end{document}